\documentclass[a4paper,12pt]{article}
\pdfoutput=1 

\usepackage{jheppub_X} 

\usepackage[T1]{fontenc} 
\usepackage{amsmath}
\usepackage{dsfont}
\usepackage{slashed}
\usepackage{bm}
\usepackage{enumerate}

\usepackage{longtable}
\usepackage{multirow}

\usepackage{hhline}
\usepackage{rotating}
\usepackage{array}
\usepackage{float}
\usepackage{afterpage}
\usepackage{anyfontsize}
\usepackage{enumitem}
\usepackage[dvipsnames,table]{xcolor}
\usepackage{arydshln}
\usepackage{blkarray}
\usepackage{physics}
\usepackage{cleveref}
\usepackage{mathtools}
\usepackage{xspace}
\usepackage{setspace}
\usepackage[normalem]{ulem}

\usepackage[most]{tcolorbox}
\definecolor{light-gray}{gray}{0.9}
\usepackage{bbold}

\makeatletter
\g@addto@macro\bfseries{\boldmath}
\makeatother

\crefname{table}{Table}{Tables}
\crefname{equation}{Eq.}{Eqs.}
\crefname{appendix}{App.}{Apps.}
\crefname{section}{Sec.}{Secs.}
\crefname{figure}{Fig.}{Figs.}

  {}

\def\L{{\cal L}}
\def\O{{\cal O}}
\def\Dcal{{\cal D}}
\def\Acal{{\cal A}}

\def\ie{\emph{i.e.}}
\def\eg{\emph{e.g.}}

\def\mat#1{\boldsymbol{#1}}

\def\T{\textsf{T}}
\def\X{\mat{\widehat{P}}}
\def\P{\widehat{P}}
\def\G5{\mat{\Gamma}^5}
\def\F{F}

\def\Dsl{\slashed{D}}

\def\Gsl{\slashed{G}}
\def\pdsl{\slashed\partial}
\def\eps{\varepsilon}
\def\Vsl{\slashed{V}}

\def\alphamat{\mat{\alpha}}
\def\btr{\mathbf{tr}}

\newcommand{\s}{\hspace{0.8pt}}

\allowdisplaybreaks

\preprint{
\vspace{-8pt}
\begin{flushright}CERN-TH-2023-002\end{flushright}
}

\title{\Large Anomaly Cancellation in Effective Field Theories \\ From the Covariant Derivative Expansion}

\author[1,2,3]{Timothy~Cohen,}
\author[4,3]{Xiaochuan~Lu,}
\author[5]{and Zhengkang~Zhang}

\affiliation[1]{\fontsize{9}{9}\selectfont \,Theoretical Physics Department, CERN, 1211 Geneva, Switzerland}
\affiliation[2]{\fontsize{9}{9}\selectfont \,Theoretical Particle Physics Laboratory, EPFL, 1015 Lausanne, Switzerland}
\affiliation[3]{\fontsize{9}{9}\selectfont \,Institute for Fundamental Science, University of Oregon, Eugene, OR 97403, USA}
\affiliation[4]{\fontsize{9}{9}\selectfont \,Department of Physics, University of California, San Diego, La Jolla, CA 92093, USA}
\affiliation[5]{\fontsize{9}{9}\selectfont \,Department of Physics, University of California, Santa Barbara, CA 91106, USA}

\emailAdd{tim.cohen@cern.ch}
\emailAdd{xil224@ucsd.edu}
\emailAdd{zkzhang@ucsb.edu}

\abstract{We extend our recently-proposed formalism for calculating anomalies of global and gauge symmetries using the Covariant Derivative Expansion to include a general class of operators that can appear in relativistic Effective Field Theories (EFTs). This allows us to prove that EFT operators involving general scalar, vector, and tensor couplings to fermion bilinears only give rise to irrelevant anomalies, which can be removed by an appropriate choice of counterterms, thereby confirming the absence of new constraints from anomaly cancellation on the Standard Model EFT.}

\begin{document}
\maketitle
\flushbottom
\setcounter{page}{2}

\begin{spacing}{1.1}
\parskip=0ex
\vspace{6pt}
\section{Introduction}
\label{sec:Introduction}

Anomalies provide critical consistency conditions on gauge theories such as the Standard Model; see \eg\ Ref.~\cite{Bertlmann:1996xk,Bilal:2008qx} for reviews. Anomaly cancellation in the Standard Model itself is of course well understood.  However, anomaly cancellation for Effective Field Theories (EFTs) with higher-dimensional operators is a more subtle issue, which has received renewed interest recently in the context of the Standard Model Effective Field Theory (SMEFT)~\cite{Bonnefoy:2020tyv,Feruglio:2020kfq} (see also Refs.~\cite{Marculescu:1978pv,Manohar:1984uq,Minn:1986ba,Kim:1988xy,Soto:1990ij,Dixon:1991wu} for earlier studies). As shown in these papers, demonstrating anomaly cancellation for SMEFT involves carefully accounting for the interplay of various interactions encoded in the higher-dimensional operators.  

In this paper, we generalize the method developed in Ref.~\cite{Paper1} for computing anomalies with the Covariant Derivative Expansion (CDE)~\cite{Gaillard:1985uh, Chan:1986jq, Cheyette:1987qz, Henning:2014wua, Henning:2016lyp} to the case of EFTs. This will allow us to confirm that the anomaly cancellation condition is unchanged by the presence of a general class of higher-dimensional operators.  More precisely, contributions to anomalies from higher-dimensional operators are in the form of the gauge variation of local operators. These are known as irrelevant anomalies and can be removed by the renormalization procedure with appropriate counterterms (see \eg\ Ref.~\cite{Cornella:2022hkc} for a recent systematic study of such counterterms focused on renormalizable theories). In contrast, relevant anomalies are IR effects and are not affected by higher-dimensional operators. We will demonstrate this explicitly with a CDE calculation.

To see that anomalies \emph{a priori} may depend on the detailed form of the interactions in the theory, let us briefly review its definition. We extract anomalies from the gauge variation of the bosonic effective action $W[G^\mu]$, defined as
\begin{equation}
e^{iW[G^\mu]} \equiv \int \Dcal\chi\s \Dcal\chi^\dagger \, e^{iS[\chi, \chi^\dagger, G^\mu]} \,.
\end{equation}
Even when the classical action is gauge invariant, $S[\chi_\alpha, \chi_\alpha^\dagger, G_\alpha^\mu] = S[\chi, \chi^\dagger, G^\mu]$ (where subscript $\alpha$ denotes the gauge-transformed quantity), the bosonic effective action after integrating out the fermions may not be:
\begin{equation}
W[G_\alpha^\mu] \stackrel{?}{=} W[G^\mu] \,.
\end{equation}
This possible discrepancy is due to the path integral measure $\Dcal\chi \Dcal\chi^\dagger$:
\begin{align}
e^{iW[G_\alpha^\mu]} &= \int \Dcal\chi\s \Dcal\chi^\dagger \, e^{iS[\chi, \chi^\dagger, G_\alpha^\mu]}
= \int \Dcal\chi_\alpha \Dcal\chi_\alpha^\dagger \, e^{iS[\chi_\alpha, \chi_\alpha^\dagger, G_\alpha^\mu]}
\notag\\[5pt]
&= \int \mathcal{J}_\alpha^{-1} \Dcal\chi\s \Dcal\chi^\dagger \, e^{iS[\chi, \chi^\dagger, G^\mu]}
= e^{iW[G^\mu]} \left\langle \mathcal{J}_\alpha^{-1} \right\rangle_G \,,
\end{align}
or equivalently
\begin{equation}
W[G_\alpha^\mu] - W[G^\mu] = -i\log \left\langle \mathcal{J}_\alpha^{-1}  \right\rangle_G = \Acal[\alpha]
+ \O(\alpha^2) \,.
\label{eqn:defineanomaly}
\end{equation}
We see that the anomaly functional $\Acal[\alpha]$ (which we will often just refer to as the anomaly), as defined by the first-order gauge variation of the bosonic effective action, is related to the expectation value of the Jacobian factor $\left\langle \mathcal{J}_\alpha^{-1} \right\rangle_G$. As emphasized by the subscript $G$, this expectation value may \emph{a priori} depend on the details of the theory, namely what interactions it contains (just as expectation values of generic operators would). So one needs a general formalism to calculate the anomalies for theories with generic interactions. 

In Ref.~\cite{Paper1}, we focused on the case of chiral fermions minimally coupled to gauge fields and introduced a regularization prescription -- a generalized version of the classic Fujikawa's method~\cite{Fujikawa:1979ay,Fujikawa:1980eg,Fujikawa:1983bg,Fujikawa:2004cx} -- to efficiently evaluate the anomaly in $d=4$ spacetime dimensions using CDE. This approach leads to unambiguous evaluation results, in the form of a master formula for the anomaly functional $\Acal[\alpha]$ that integrates various known results about anomalies. In this paper, we extend this formalism to include a more general set of interactions in Lorentz-invariant EFTs such as SMEFT.

The rest of this paper is organized as follows. In \cref{sec:Parameterization} we present our parameterization of a general class of EFT operators, involving scalar, vector, and tensor couplings to fermion bilinears. In \cref{sec:CDE} we generalize the formalism in Ref.~\cite{Paper1} and explain how to calculate the anomaly in such EFTs with CDE. We complete the detailed evaluation of the anomaly in \cref{sec:Irrelevance} and show that extra contributions from the interactions beyond minimal coupling are all irrelevant anomalies. Finally, in \cref{sec:Discussion} we conclude and discuss some future directions.

\section{Parameterization of a General EFT}
\label{sec:Parameterization}

We are interested in anomalies of both gauge and global symmetries in a general Lorentz-invariant EFT. As in Ref.~\cite{Paper1}, we introduce \emph{auxiliary} gauge fields for all the global symmetries of interest. Putting them together with the \emph{physical} gauge fields, we denote the whole collection by $G_\mu$, which can be a sum over multiple (Abelian and/or non-Abelian) group sectors:
\begin{equation}
G_\mu \equiv \sum_{a} G_\mu^a t^a \,.
\label{eqn:Gdef}
\end{equation}
The (Hermitian) covariant derivative is
\begin{equation}
P_\mu \equiv i D_\mu = i \partial_\mu + G_\mu \,,
\end{equation}
and the gauge field strength is given by
\begin{equation}
\F_{\mu\nu} = \sum_a \F_{\mu\nu}^a\, t^a = -i\, \comm{P_\mu}{P_\nu} = \left(\partial_\mu G_\nu\right) - \left(\partial_\nu G_\mu\right) - i\, \comm{G_\mu}{G_\nu} \,.
\end{equation}
We consider a general theory of $n$ left-handed Weyl fermions $\chi_1, \dots, \chi_n$, with each $\chi_i$ transforming in an irreducible representation of the (global and gauge) symmetries. The theory may also contain an arbitrary number of scalar fields, collectively denoted as $\phi$. The EFT Lagrangian we consider has the following general form:
\begin{align}
\L =&\; \L_{G,\phi} + \sum_{i=1}^n \chi^\dagger_i \,\overline{\sigma}^\mu P_\mu\, \chi_i \notag\\[3pt]
& + \sum_{i,j=1}^n \bigg\{ \chi^\dagger_i\,\overline{\sigma}_\mu V^\mu_{ij}\, \chi_j + \Big[ \chi_i \bigl(S_{ij} + i\sigma_\mu\overline{\sigma}_\nu T^{\mu\nu}_{ij}\bigr) \chi_j + \text{h.c.} \Big] \bigg\} \,.
\label{eq:L_explicit}
\end{align}
Here $\L_{G,\phi}$ collects the interactions that do not involve fermions. The rest of the first line encodes the minimal couplings between the fermions $\chi_i$ and gauge fields $G_\mu$. In the second line, we parameterize an extended set of interactions with fermion bilinears, categorizing them into scalar, vector, and tensor interactions:
\begin{equation}
\chi_i\, S_{ij}\, \chi_j \,,\qquad
\chi_i^\dagger\, \overline\sigma_\mu V_{ij}^\mu\, \chi_j \,,\qquad
\chi_i\, i\sigma_\mu \overline\sigma_\nu T_{ij}^{\mu\nu}\, \chi_j \,,
\end{equation}
where $S_{ij}[G,\phi]$, $V_{ij}^\mu[G,\phi]$, $T_{ij}^{\mu\nu}[G,\phi]$ are functions made of $G_\mu$, $\phi$ and their derivatives and can have arbitrarily high operator dimensions. Note that due to the Clifford algebra $\sigma_\mu \overline\sigma_\nu + \sigma_\nu \overline\sigma_\mu = 2\eta_{\mu\nu}\mathbb{1}$, the $\mu\leftrightarrow\nu$ symmetric components of $T_{ij}^{\mu\nu}$ can be absorbed into the scalar interactions $S_{ij}$, so we define the tensor interactions to be antisymmetric, $T_{ij}^{\nu\mu}=-T_{ij}^{\mu\nu}$.

In the equations above, we have been using the standard two-component notation and have suppressed the spinor indices. Taking the scalar interactions for example, if we write out the spinor indices and put the expression into matrix form, we have
\begin{equation}
\chi_i \chi_j = (\chi_i)^\alpha (\chi_j)_\alpha = (\chi_i)_\beta \left(-\epsilon^{\beta\alpha}\right) (\chi_j)_\alpha
\quad\longrightarrow\quad
\chi_i^\T \left( -i\sigma^2 \right) \chi_j \,,
\end{equation}
which is symmetric under $i \leftrightarrow j$. Absorbing the $i,j$ indices also into matrix form, we can write
\begin{subequations}\label{eqn:SVTmat}
\begin{align}
\chi_i\, S_{ij}\, \chi_j
&\quad\longrightarrow\quad
\chi^\T \left( -i\sigma^2 S \right) \chi \,, \\[5pt]
\chi_i^\dagger\, \overline\sigma_\mu V_{ij}^\mu\, \chi_j
&\quad\longrightarrow\quad
\chi^\dagger \left( \overline\sigma^\mu V_\mu \right) \chi \,, \\[5pt]
\chi_i\, i\sigma_\mu \overline\sigma_\nu T_{ij}^{\mu\nu}\, \chi_j
&\quad\longrightarrow\quad
\chi^\T \left( -i\sigma^2 i\sigma^\mu \overline\sigma^\nu T_{\mu\nu} \right) \chi \,.
\end{align}
\end{subequations}
We see that without loss of generality we can require
\begin{subequations}
\begin{alignat}{3}
\left( -i\sigma^2 S \right)^\T &= - \left( -i\sigma^2 S \right)
&&\quad\Longrightarrow\quad &
S^\T &= S \,, \\[5pt]
\left( -i\sigma^2 i\sigma^\mu \overline\sigma^\nu T_{\mu\nu} \right)^\T &= - \left( -i\sigma^2 i\sigma^\mu \overline\sigma^\nu T_{\mu\nu} \right)
&&\quad\Longrightarrow\quad &
T_{\mu\nu}^\T &= T_{\nu\mu} = - T_{\mu\nu} \,.
\end{alignat}
\end{subequations}
Furthermore, Hermiticity of the Lagrangian in \cref{eq:L_explicit} requires $V_\mu^\dagger = V_\mu$.

A general symmetry transformation of the fermions can be parameterized as
\begin{equation}
\chi
\quad\rightarrow\quad
\chi_\alpha \equiv U_\alpha\, \chi = e^{i\alpha}\, \chi \,,
\label{eqn:GTfermions}
\end{equation}
where, similar to the gauge fields in \cref{eqn:Gdef}, $\alpha \equiv \alpha^a t^a$ is a sum over all symmetry group generators (across multiple sectors). For the Lagrangian in \cref{eq:L_explicit} to respect the (global and gauge) symmetries, we need the following transformation properties of various quantities:
\begin{subequations}\label{eqn:GTvarious}
\begin{align}
P^\mu
&\quad\longrightarrow\quad
P^\mu_\alpha = U_\alpha\, P^\mu\, U_\alpha^\dagger \,,\quad
&\delta_\alpha P^\mu &= \delta_\alpha G^\mu = i\comm{\alpha}{P^\mu} \,, \\[5pt]
V^\mu
&\quad\longrightarrow\quad
V^\mu_\alpha = U_\alpha\, V^\mu\, U_\alpha^\dagger \,,\quad
&\delta_\alpha V^\mu &= i\comm{\alpha}{V^\mu} \,, \\[5pt]
S
&\quad\longrightarrow\quad
S_\alpha = U_\alpha^*\, S\, U_\alpha^\dagger \,,\quad
&\delta_\alpha S &= - i(\alpha^\T S + S \alpha) \,, \\[5pt]
T^{\mu\nu}
&\quad\longrightarrow\quad
T^{\mu\nu}_\alpha = U_\alpha^*\, T^{\mu\nu}\, U_\alpha^\dagger \,,\quad
&\delta_\alpha T^{\mu\nu} &= - i(\alpha^\T T^{\mu\nu} + T^{\mu\nu} \alpha) \,, \\[5pt]
\L_{G,\phi}
&\quad\longrightarrow\quad
\L_{G,\phi} [G_\alpha, \phi_\alpha] = \L_{G,\phi}[G, \phi] \,,\quad
&\delta_\alpha \L_{G,\phi} &= 0 \,,
\end{align}
\end{subequations}
where $\delta_\alpha$ denotes the first order (in $\alpha$) gauge variation, \eg\ $\delta_\alpha P^\mu \equiv (P^\mu_\alpha - P^\mu)|_{\O(\alpha)}$.

The Lagrangian in \cref{eq:L_explicit} incorporates the most general scalar, vector, and tensor couplings to fermion bilinears. For example, in SMEFT, the vector interactions $V_{ij}^\mu$ cover current-current operators such as
\begin{equation}
\bigl(H^\dagger i\overleftrightarrow{D}_\mu H\bigr) \bigl(\bar\psi \gamma^\mu \psi \bigr) \,,\quad
|H|^2 \bigl(H^\dagger i\overleftrightarrow{D}_\mu H\bigr) \bigl(\bar\psi \gamma^\mu \psi \bigr) \,,\quad
\cdots
\end{equation}
where $H$ is the Higgs doublet and $\psi$ may represent any of the SM fermions $\psi \in \{q, u, d, \ell, e\}$, written in four-component notation here. The scalar interactions $S_{ij}$ cover Yukawa-type operators such as
\begin{equation}
{\bar \ell} e H \,,\quad
|H|^2\, {\bar \ell} e H \,,\quad
\cdots
\label{eq:S_operators}
\end{equation}
The tensor interactions $T_{ij}^{\mu\nu}$ cover dipole operators like
\begin{equation}
{\bar \ell} \sigma^{\mu\nu} e B_{\mu\nu} H \,,\quad
|H|^2\, {\bar \ell} \sigma^{\mu\nu} e B_{\mu\nu} H \,,\quad
\cdots
\end{equation}
where $\sigma^{\mu\nu}\equiv \frac{i}{2}[\gamma^\mu, \gamma^\nu]$. At dimension six, these include all but the four-fermion operators in the Warsaw basis~\cite{Grzadkowski:2010es}. In fact, all operators in SMEFT involving up to two powers of fermions and no derivatives acting on them, up to arbitrarily high dimensions, are captured by \cref{eq:L_explicit}. Furthermore, as argued in Refs.~\cite{Soto:1990ij,Feruglio:2020kfq}, four-fermion operators can be captured by introducing auxiliary fields, and we believe this argument may be extended to operators with six and more fermions by perturbatively including interactions among such auxiliary fields. Therefore, our calculation in what follows should apply quite generally to all higher-dimensional SMEFT (as well as other relativistic EFT) operators\footnote{Throughout this paper we use the term `higher-dimensional operators' to emphasize the relevance of our analysis for the infinite series of EFT operators, though technically the operators considered here include also renormalizable ones like dimension-four Yukawa interactions (as part of $S$ -- see \cref{eq:S_operators}).} with no derivatives acting on the fermions.

\section{Regularizing and Evaluating the Anomaly with CDE}
\label{sec:CDE}

To facilitate the calculation of anomalies, we first recast the fermionic interactions in \cref{eq:L_explicit} into the following matrix form
\begin{equation}
\L = \L_{G, \phi} + \frac12 \, \begin{pmatrix} \chi^\dagger & \chi^\T \left(-i\sigma^2\right) \end{pmatrix} \X
\begin{pmatrix} \chi \\ i\sigma^2 \chi^* \end{pmatrix} \,,
\label{eq:L_matrix}
\end{equation}
where
\begin{equation}
\X
\equiv \begin{pmatrix}
\;\; \overline\sigma^\mu (i\partial_\mu +G_\mu + V_\mu) \;\;& S^\dagger -i\,\overline{\sigma}^\nu \sigma^\mu\, T^\dagger_{\mu\nu} \\[3pt]
S +i \sigma^\mu \overline{\sigma}^\nu\, T_{\mu\nu} &\;\; \sigma^\mu (i\partial_\mu -G^\T_\mu -V^\T_\mu) \;\;
\end{pmatrix} \,.
\label{eqn:Xdef}
\end{equation}
For the Lagrangian to be real and symmetry preserving, the matrix $\X$ needs to be hermitian, $\X^\dagger = \X$, and transform as
\begin{equation}
\X
\quad\longrightarrow\quad
\X_\alpha = e^{i\alphamat}\, \X \, e^{-i\alphamat} \,,\qquad
\delta_\alpha \X = i \bigl[ \alphamat, \X \bigr] 
\quad\text{with}\quad
\alphamat \equiv \Bigl(\begin{smallmatrix} \alpha & 0 \\[2pt] 0 & -\alpha^\T \end{smallmatrix}\Bigr) \,.
\label{eqn:GTX}
\end{equation}
Clearly, these properties of $\X$ can be verified using its explicit expression in \cref{eqn:Xdef} together with the transformation properties given in \cref{eqn:GTvarious}.

As reviewed in Ref.~\cite{Paper1} (see also Refs~\cite{Bertlmann:1996xk,Bilal:2008qx}), anomalies can be derived from the gauge variation of the bosonic effective action obtained by integrating out the fermion fields. In the present case, the bosonic effective action depends on both gauge and scalar fields (\ie, we integrate out the fermions in the path integral while treating all bosonic fields as classical backgrounds). Formally, we have
\begin{equation}
e^{iW[G, \phi]} = \int \Dcal \chi \Dcal \chi^\dagger\, e^{iS[\chi, \chi^\dagger, G, \phi]}
= e^{iS_{G,\phi}} \bigl(\det\X\bigr)^{1/2} \,.
\end{equation}
Following \cref{eqn:GTX}, a gauge transformation yields
\begin{equation}
e^{iW[G_\alpha, \phi_\alpha]}
= e^{iS_{G,\phi}} \bigl(\det \X_\alpha \bigr)^{1/2}
= e^{iS_{G,\phi}} \left[ \det\left( e^{i\alphamat}\, \X \, e^{-i\alphamat} \right) \right]^{1/2} \,.
\end{equation}
As in \cref{eqn:defineanomaly}, the anomaly functional (or simply the anomaly) is defined by the first-order gauge variation of the bosonic effective action:
\begin{align}
\Acal [\alpha] &\equiv \delta_\alpha W[G,\phi]
=\bigl( W[G_\alpha, \phi_\alpha] - W[G,\phi] \bigr) \Bigr|_{\O(\alpha)}
\notag\\[8pt]
& \simeq -\frac{i}{2} \Tr\log \bigg[ \mat{1} + \frac{1}{\X} \,\Bigl(i\alphamat \X - \X i\alphamat \Bigr) \bigg] \biggr|_{\O(\alpha)}
\simeq \frac12 \Tr \biggl[ \frac{1}{\X} \,\Bigl(\alphamat \X - \X \alphamat \Bigr) \biggr] \,.
\label{eq:anomaly_unregularized}
\end{align}
As explained in detail in Ref.~\cite{Paper1}, we use the notation `$\simeq$' to emphasize that the expressions are not exactly equal unless they are regularized in the same way.

To proceed further, we need to introduce a regulator. Following similar steps as in Sec.~3 of Ref.~\cite{Paper1}, we see that in the present case, each term in the expansion is proportional to
\begin{small}
\begin{equation}
\tr \mqty(
\sigma^{\mu_1} \overline{\sigma}^{\nu_1} \dots \sigma^{\mu_n} \overline{\sigma}^{\nu_n} & \cdot \\
\cdot & \overline{\sigma}^{\mu_1} \sigma^{\nu_1} \dots \overline{\sigma}^{\mu_n} \sigma^{\nu_n})
= \tr \Bigg[\gamma^{\mu_1}\gamma^{\nu_1} \dots \gamma^{\mu_n}\gamma^{\nu_n} 
\begin{pmatrix}
\tfrac{1-\gamma^5}{2} & \cdot \\
\cdot & \tfrac{1+\gamma^5}{2}
\end{pmatrix} \Bigg] \,.
\end{equation}
\end{small}
We can therefore replace all the $2\times2$ Pauli matrices by $4\times4$ gamma matrices while freely inserting chirality projection factors $\frac{1\mp\gamma^5}{2}+\beta\,\frac{1\pm\gamma^5}{2}$,  such that terms proportional to $\beta$ will hit the opposite chirality projection operator when anti-commuted to the right and vanish. To this end, let us define
\begin{equation}
\X_\beta \equiv
\left(
\begin{smallmatrix}
	i\slashed{\partial} +\slashed{G} \bigl(\frac{1-\gamma^5}{2} + \beta_G\, \frac{1+\gamma^5}{2}\bigr) +\slashed{V} \,\bigl(\frac{1-\gamma^5}{2} + \beta_V \frac{1+\gamma^5}{2}\bigr) &
	S^\dagger\bigl(\frac{1+\gamma^5}{2} + \beta_S \frac{1-\gamma^5}{2}\bigr) +\sigma^{\mu\nu} T^\dagger_{\mu\nu} \bigl(\frac{1+\gamma^5}{2} + \beta_T \frac{1-\gamma^5}{2}\bigr) \\[4pt]
	S\bigl(\frac{1-\gamma^5}{2} + \beta_S \frac{1+\gamma^5}{2}\bigr) +\sigma^{\mu\nu} T_{\mu\nu} \bigl(\frac{1-\gamma^5}{2} + \beta_T \frac{1+\gamma^5}{2}\bigr) & 
	i\slashed{\partial} -\slashed{G}^\T \bigl(\frac{1+\gamma^5}{2} + \beta_G\, \frac{1-\gamma^5}{2}\bigr) -\slashed{V}^\T \,\bigl(\frac{1+\gamma^5}{2} + \beta_V \frac{1-\gamma^5}{2}\bigr)
\end{smallmatrix}\right) \,,
\label{eq:X_beta}
\end{equation}
We clarify that $\slashed{G}^\T$, $\slashed{V}^\T$ are defined as
\begin{equation}
\slashed{G}^\T \equiv \gamma^\mu G_\mu^\T \,,\qquad
\qquad
\slashed{V}^\T \equiv \gamma^\mu V_\mu^\T \,,
\label{eqn:GVslT}
\end{equation}
in which the gamma matrices are \emph{not} transposed. Here $\beta_G, \beta_S, \beta_V, \beta_T$ can all be different in principle, and we denote them collectively as $\beta$ in the subscript of $\X_\beta$ (and also $\Acal_\beta$ below). Replacing $\X \to \X_\beta$ in \cref{eq:anomaly_unregularized}, we see that similarly to Ref.~\cite{Paper1}, a damping factor $f\bigl(-\X_\beta^2/\Lambda^2\bigr)$ emerges as a natural regulator, with any function $f(u)$ that satisfies the following conditions
\begin{subequations}\label{eqn:fconditions}
\begin{align}
f(0) = 1 \,;\qquad
f(+\infty) = 0 \,;\qquad
\int_0^\infty  \dd u\s f(u)
&\quad\text{well defined}\,, \\[10pt]
\left. u^n \frac{\dd^n f}{\dd u^n} \,\right|_{u=0} = \left. u^n \frac{\dd^n f}{\dd u^n}\, \right|_{u\to+\infty} = 0
&\quad\text{for}\quad
n\ge1 \,.
\end{align}
\end{subequations}
The \emph{regularized} anomaly is then defined by
\begin{equation}
\Acal_\beta^\Lambda[\alpha] \equiv \frac12 \Tr \left[ f\Bigl(-\tfrac{\X_\beta^2}{\Lambda^2}\Bigr)\, \X_\beta^{-1} \left( \alphamat \X_\beta - \X_\beta \alphamat \right) \frac{1-\G5}{2} \right] \,,
\end{equation}
where we have introduced the notation
\begin{equation}
\G5 \equiv \left(\begin{smallmatrix} \gamma^5 & 0 \\[2pt] 0 & -\gamma^5 \end{smallmatrix}\right) 
\qquad\text{satisfying}\qquad
\X_\beta \G5 = - \G5 \X_\beta \,.
\end{equation}
Now using the cyclicity of the trace and commuting $\X_\beta$ through $f\Bigl(-\tfrac{\X_\beta^2}{\Lambda^2}\Bigr)$, we obtain
\begin{equation}
\Acal_\beta^\Lambda [\alpha] = \frac12 \Tr \biggl[ f\Bigl( -\tfrac{\X_\beta^2}{\Lambda^2} \Bigr) \, \G5\, \alphamat \biggr] \,.
\label{eq:anomaly_beta}
\end{equation}
The \emph{renormalized} anomaly is then defined by
\begin{equation}
\Acal_\beta[\alpha] \equiv \lim_{\Lambda\to\infty} \Bigl( \Acal_\beta^\Lambda[\alpha] +\delta_\alpha \int \dd^4x \,\L_\text{ct}^\Lambda \Bigr) \,,
\label{eq:A_ren}
\end{equation}
where $\L_\text{ct}^\Lambda$ is the local counterterm Lagrangian. Since $\Acal_\beta^\Lambda[\alpha]$ may be quadratically divergent, we must include appropriate $\O(\Lambda^2)$ counterterms to make the renormalized anomaly finite in the limit $\Lambda\to\infty$. Meanwhile, the finite part of $\L_\text{ct}^\Lambda$ defines the renormalization scheme. Generically $\Acal_\beta^\Lambda[\alpha]$ also contains $\O(1/\Lambda)$ terms, which we will suppress throughout the paper since they vanish when $\Lambda\to\infty$.

\cref{eq:anomaly_beta} is a generalization of the minimal coupling (mc) case formula in Ref.~\cite{Paper1}. To see the connection explicitly, we note that when $S = V_\mu = T_{\mu\nu}=0$, $\X_\beta$ becomes block diagonal with the two blocks related by charge conjugation:\footnote{Note that $(\partial_\mu)^\T = \overleftarrow{\partial}_\mu = -\partial_\mu$ upon integration by parts.}
\begin{equation}
\X_\beta^\text{mc}
= \left(\begin{smallmatrix} \P_\beta & 0 \\[4pt] 
0 & \overline{\P}_\beta \end{smallmatrix}\right) 
= \left(\begin{smallmatrix}
i\slashed{\partial} + \gamma^\mu G_\mu \bigl(\frac{1-\gamma^5}{2} + \beta_G\, \frac{1+\gamma^5}{2}\bigr) & 0 \\[4pt]
0 & -i\slashed{\partial}^\T -\gamma^\mu G_\mu^\T \bigl(\frac{1+\gamma^5}{2} + \beta_G\, \frac{1-\gamma^5}{2}\bigr)
\end{smallmatrix}\right) \,,
\end{equation}
where `charge conjugation' is the operation
\begin{equation}
\overline{\P}_\beta \equiv \gamma^0 \gamma^2\, \P_\beta^\T\, \gamma^0 \gamma^2 \,,
\end{equation}
under which the gamma matrices transform as
\begin{equation}
\overline{\gamma^\mu} = -\gamma^\mu \,,\qquad
\overline{\sigma^{\mu\nu}} = -\sigma^{\mu\nu} \,,\qquad
\overline{\gamma^5} = \gamma^5 \,.
\end{equation}
It satisfies the expected properties:
\begin{equation}
\overline{\overline{A}} = A \,,\qquad\quad
\tr \bigl( A B C \cdots \bigr) = \tr \bigl( \cdots \overline{C}\, \overline{B}\,\overline{A} \bigr) \,.
\label{eqn:CCProperties}
\end{equation}
Therefore, the two blocks in $\X_\beta^\text{mc}$ contribute equally, and \cref{eq:anomaly_beta} reduces to the result derived in Ref.~\cite{Paper1}:
\begin{equation}
\Acal_\beta^{\Lambda,\text{mc}}[\alpha] = \Tr \biggl[ f\Bigl(-\tfrac{\P_\beta^2}{\Lambda^2}\Bigr)\, \gamma^5\, \alpha \biggr] \,.
\end{equation}

It is useful to introduce an extended version of this charge conjugation operation. For a matrix $\mat{A}$ acting on the field multiplet space, we define 
\begin{equation}
\overline{\mat{A}} \equiv
\gamma^0\gamma^2 
\Bigl(\begin{smallmatrix}
0 & 1 \\[4pt] 1 & 0
\end{smallmatrix}\Bigr) \,
\mat{A}^\T \,
\gamma^0\gamma^2 
\Bigl(\begin{smallmatrix}
0 & 1 \\[4pt] 1 & 0
\end{smallmatrix}\Bigr) \,.
\label{eqn:CCdef}
\end{equation}
Clearly, the properties in \cref{eqn:CCProperties} hold for this extended version as well, and one can also check that
\begin{equation}
\overline{\X}_\beta = \X_\beta \,,\qquad
\overline{\alphamat} = -\alphamat \,, \qquad
\overline{\G5} = -\G5 \,.
\end{equation}

The CDE evaluation of the functional trace in \cref{eq:anomaly_beta} proceeds in a similar way to the minimal coupling case detailed in Ref.~\cite{Paper1}. After performing the loop integrals, we obtain
\begin{equation}
\Acal_\beta^\Lambda [\alpha] = \frac{i}{32\pi^2} \int \dd^4x\, \bigg\{
-\Lambda^2 \left[ \int_0^\infty \dd u\s f(u) \right] \btr_0\, +\, \frac16\, \bigl(\btr_1+\btr_2 +\btr_3\bigr)
\bigg\} \,,
\end{equation}
with a few (non-functional) traces over the field multiplet and internal indices (denoted by lowercase `tr'):
\begin{subequations}\label{eq:tr}
\begin{align}
\btr_0 &= \tr \bigl( \X_\beta^2 \, \G5 \alphamat \bigr) \,, \\[8pt]
\btr_1 &= \tr \bigl( \X_\beta^4 \, \G5 \alphamat \bigr) \,, \\[8pt]
\btr_2 &= -\tfrac12 \tr \Bigl[ \bigl(\X_\beta^2\,\gamma^\mu\X_\beta\gamma_\mu \X_\beta + \X_\beta \gamma^\mu\X_\beta\gamma_\mu \X_\beta^2 \bigr) \, \G5 \alphamat \Bigr] 
\,,\\[8pt]
\btr_3 &= -\tfrac12 \tr \bigl( \X_\beta \gamma^\mu\X_\beta^2 \,\gamma_\mu \X_\beta \, \G5 \alphamat \bigr) \,.
\end{align}
\end{subequations}
These are generalizations of tr$_0$--tr$_3$ (unbolded) in Ref.~\cite{Paper1}, although (obviously) their evaluation result is twice as large, $\btr_i = 2\tr_i$,  in the minimal coupling case. It is also useful to note that the two terms in $\btr_2$ are related by charge conjugation and therefore equal,\footnote{Note that we need to use cyclic permutation inside the internal trace `$\tr$' for this argument. See App.~A of Ref.~\cite{Paper1} for a detailed discussion about the legitimacy of such operations, which will be assumed throughout this paper.} which simplifies its calculation later on.

\section{Higher-dimensional Operators Yield Irrelevant Anomalies}
\label{sec:Irrelevance}

In this section, we complete the evaluation of the regularized anomaly $\Acal_\beta^\Lambda [\alpha]$ by computing the traces in \cref{eq:tr}. For general values of $\beta_G, \beta_S, \beta_V, \beta_T$, the calculation is very tedious and does not give new insights.  The reason is that most $\beta$ choices lead to results that do not satisfy the Wess-Zumino consistency condition~\cite{Wess:1971yu}, which means they do not correspond to consistent regularization schemes of the effective action and there is no meaningful notion of relevant vs.\ irrelevant anomalies. This point has been discussed in detail in Ref.~\cite{Paper1} in the minimal coupling case where only $\beta_G$ is present; for example, $\beta_G=0$ is the only choice that satisfies the Wess-Zumino consistency condition for the case of a nontrivial non-Abelian anomaly. Motivated by the results in Ref.~\cite{Paper1}, we will set all the $\beta$'s to zero in the present analysis:
\begin{equation}
\beta_G = \beta_S = \beta_V = \beta_T = 0 \,.
\end{equation}
With this regularization scheme choice, we will show that all the additional contributions to the anomaly are irrelevant, namely:
\begin{equation}
\Acal_{\beta=0}^\Lambda[\alpha] = \Acal_{\beta=0}^{\Lambda,\text{mc}}[\alpha] - \delta_\alpha \int \dd^4x\, \Delta\L_\text{ct}^\Lambda \,.
\label{eq:irrelevant}
\end{equation}
This means that by appropriately adjusting the local counterterms (\ie\ choosing the renormalization scheme), the renormalized anomaly defined in \cref{eq:A_ren} is the same as that in the minimal coupling case:
\begin{equation}
\Acal_{\beta=0}[\alpha] = \Acal_{\beta=0}^{\text{mc}}[\alpha] \,.
\end{equation}

Setting $\beta=0$ significantly simplifies the presentation; we now have
\begin{align}
\X_{\beta=0} \equiv i\pdsl + \left(\begin{smallmatrix}
\Gsl + \Vsl & S^\dagger + \sigma^{\mu\nu} T^\dagger_{\mu\nu} \\[4pt]
 S + \sigma^{\mu\nu} T_{\mu\nu} & -\slashed{G}^\T - \slashed{V}^\T
\end{smallmatrix}\right) \frac{1-\G5}{2} \,.
\label{eqn:X0}
\end{align}
Nevertheless, the calculation including $S, V_\mu, T_{\mu\nu}$ all at once is still quite lengthy. So in what follows, we will work up to the full results gradually, adding one type of interactions at each step.

\subsection{Vector Interactions}
\label{subsec:Vector}

We begin with the case of having vector interactions $V_\mu$ only, while setting $S$ and $T_{\mu\nu}$ to zero. In this case, there is actually a shortcut. From the expression of $\X_{\beta=0}$ in \cref{eqn:X0} we see that, instead of directly calculating the traces in \cref{eq:tr}, we can simply take the minimal coupling result and replace $G_\mu\to G_\mu+V_\mu$:
\begin{equation}
\Acal_{\beta=0}^\Lambda[\alpha]\Bigr|_{S=T_{\mu\nu}=0} = \Acal_{\beta=0}^{\Lambda,\text{mc}}[\alpha] \Bigr|_{G_\mu\to G_\mu+V_\mu} \,.
\end{equation}
In Ref.~\cite{Paper1}, we obtained the result for the minimal coupling case
\begin{equation}
\Acal_{\beta=0}^{\Lambda,\text{mc}}[\alpha] = \int \dd^4x\,\biggl\{ \frac{1}{48\pi^2}\, \eps^{\mu\nu\rho\sigma} \tr
\Bigl[ \left( \partial_\mu \alpha \right) \left( G_\nu \F_{\rho\sigma} + i G_\nu G_\rho G_\sigma \right) \Bigr]
- \delta_\alpha \L_{\text{ct},0}^\Lambda \biggr\} \,.
\label{eq:anomaly_mc}
\end{equation}
The first term is the standard result for the consistent anomaly. The second term, being the gauge variation of a local counterterm
\begin{align}
\L_{\text{ct},0}^\Lambda &= \frac{1}{16\pi^2} \left[ \Lambda^2 \int_0^\infty \dd u\s f(u)\right] \tr\bigl(G^\mu G_\mu\bigr)
\notag\\[5pt]
&\hspace{20pt}
+\frac{1}{96\pi^2} \tr\biggl[ (\partial^\mu G_\mu)^2-2i\,\F^{\mu\nu} G_\mu G_\nu +\frac{1}{2} \,G^\mu G^\nu G_\mu G_\nu \biggr] \,,
\label{eq:Lct0}
\end{align}
is an irrelevant anomaly.

Upon making the substitution $G_\mu\to G_\mu+V_\mu$, we first note that the irrelevant term in \cref{eq:anomaly_mc} remains irrelevant, because the two operations `taking the gauge variation' and `substituting $G_\mu\to G_\mu+V_\mu$' commute with each other:
\begin{equation}
\left( \delta_\alpha \L_{\text{ct},0}^\Lambda \right) \Bigr|_{G_\mu\to G_\mu+V_\mu}
= \delta_\alpha \Bigl( \L_{\text{ct},0}^\Lambda \Bigr|_{G_\mu\to G_\mu+V_\mu} \Bigr) \,,
\end{equation}
due to the fact
\begin{equation}
\delta_\alpha (G_\mu + V_\mu) = (\partial_\mu\alpha) + i\,\comm{\alpha}{G_\mu} +i\, \comm{\alpha}{V_\mu}  = (\partial_\mu\alpha) +i \,\comm{\alpha}{G_\mu + V_\mu} \,.
\end{equation}

For the relevant part of $\Acal_{\beta=0}^{\Lambda,\text{mc}}$ (first term in \cref{eq:anomaly_mc}), the substitution $G_\mu\to G_\mu+V_\mu$ produces additional terms that we need to track carefully. Using
\begin{equation}
\F_{\mu\nu} \bigr|_{G_\mu\to G_\mu+V_\mu} =
\F_{\mu\nu} + (D_\mu V_\nu) - (D_\nu V_\mu) - i\,\comm{V_\mu}{V_\nu} \,,
\end{equation}
where $(D_\mu V_\nu) \equiv (\partial_\mu V_\nu) - i\,\comm{G_\mu}{V_\nu}$, we get
\begin{align}
\Acal_{\beta=0}^\Lambda[\alpha]\bigr|_{S=T_{\mu\nu}=0} =\;& \Acal_{\beta=0}^{\Lambda,\text{mc}}[\alpha] - \delta_\alpha \int \dd^4x \left( \L_{\text{ct},0}^\Lambda\Bigr|_{G_\mu\to G_\mu+V_\mu} - \L_{\text{ct},0}^\Lambda \right)
\notag\\[5pt]
&\hspace{0pt}
- \int \dd^4x\, \frac{1}{48\pi^2}\, \eps^{\mu\nu\rho\sigma} \tr\bigg\{
(\partial_\mu \alpha) \Big[
- \left( V_\nu G_{\rho\sigma} + G_{\rho\sigma} V_\nu \right)
\notag\\[5pt]
&\hspace{30pt}
- i \left( G_\nu G_\rho V_\sigma + V_\nu G_\rho G_\sigma - G_\nu V_\rho G_\sigma \right)
- 2 V_\nu \left(D_\rho V_\sigma\right)
\notag\\[5pt]
&\hspace{30pt}
- i V_\nu G_\rho V_\sigma + i \left( G_\nu V_\rho V_\sigma - V_\nu V_\rho G_\sigma \right)
+ i V_\nu V_\rho V_\sigma
\Big] \bigg\} \,.
\label{eqn:AcalVshift}
\end{align}
Using the gauge transformation properties of the various quantities:
\begin{subequations}\label{eqn:GTGV}
\begin{alignat}{2}
\delta_\alpha G_\mu &= (\partial_\mu \alpha) + i\,\comm{\alpha}{G_\mu} \,,&\qquad\quad
\delta_\alpha G_{\mu\nu} &= i\,\comm{\alpha}{G_{\mu\nu}} \,, \\[5pt]
\delta_\alpha V_\mu &= i\,\comm{\alpha}{V_\mu} \,,&\qquad
\delta_\alpha (D_\mu V_\nu) &= i\,\comm{\alpha}{(D_\mu V_\nu)} \,,
\end{alignat}
\end{subequations}
we can organize the terms beyond the first line in \cref{eqn:AcalVshift} into the gauge variation of the following local counterterm:
\begin{align}
\Delta\L_\text{ct}^{(V)} &= \frac{1}{48\pi^2}\, \eps^{\mu\nu\rho\sigma} \tr\bigg[
-G_\mu \left( V_\nu \F_{\rho\sigma} + \F_{\rho\sigma} V_\nu \right) -i G_\mu G_\nu G_\rho V_\sigma
- 2 G_\mu V_\nu \left(D_\rho V_\sigma\right)
\notag\\[5pt]
&\hspace{123pt}
- \frac{i}{2}\, G_\mu V_\nu G_\rho V_\sigma + i G_\mu G_\nu V_\rho V_\sigma
+ i G_\mu V_\nu V_\rho V_\sigma
\bigg] \,.
\label{eq:LctV}
\end{align}
Note that when taking the gauge variation of the expression above, all the commutator terms generated through \cref{eqn:GTGV} cancel out, which leaves us only with terms proportional to $(\partial_\mu \alpha)$, reproducing the expression in \cref{eqn:AcalVshift}. In summary, we have shown that
\begin{equation}
\Acal_{\beta=0}^\Lambda [\alpha]\Bigr|_{S=T_{\mu\nu}=0} = \Acal_{\beta=0}^{\Lambda,\text{mc}}[\alpha] - \delta_\alpha \int \dd^4x \left( \L_{\text{ct},0}^\Lambda\Bigr|_{G_\mu\to G_\mu+V_\mu} - \L_{\text{ct},0}^\Lambda + \Delta\L_{\text{ct}}^{(V)} \right) .
\end{equation}
We conclude that all additional contributions to the anomaly due to the vector interactions $V_\mu$ are irrelevant.

\subsection{Vector and Scalar Interactions}
\label{subsec:Scalar}

In this subsection, we turn on both the scalar interactions $S$ and vector interactions $V_\mu$ while keeping $T_{\mu\nu}=0$. We will further include the tensor interactions $T_{\mu\nu}$ in the next subsection.

To calculate the traces in \cref{eq:tr} in the presence of $S$ and/or $T_{\mu\nu}$, it is useful to decompose $\X_{\beta=0}$ in \cref{eqn:X0} as
\begin{align}
\X_{\beta=0} &= \left( S_L + \gamma_\mu V_L^\mu + \sigma_{\mu\nu} T_L^{\mu\nu} \right) \frac{1-\gamma^5}{2}
+ \left( S_R + \gamma_\mu V_R^\mu + \sigma_{\mu\nu} T_R^{\mu\nu} \right) \frac{1+\gamma^5}{2} \notag\\[5pt]
&\equiv \left( S_L + V_L + T_L \right) \frac{1-\gamma^5}{2} + \left( S_R + V_R + T_R \right) \frac{1+\gamma^5}{2} \,,
\label{eqn:X0decompose}
\end{align} 
where we have introduced the notation:
\begin{subequations}\label{eqn:chiralVST}
\begin{alignat}{3}
V_L &= \Bigl(\begin{smallmatrix} i\pdsl + \Gsl + \Vsl & 0 \\[4pt]
            0 & \quad i\pdsl \quad \end{smallmatrix}\Bigr) \,,&\qquad
S_L &= \Bigl(\begin{smallmatrix} 0 & 0 \\[4pt] S & 0 \end{smallmatrix}\Bigr) \,,&\qquad
T_L &= \sigma_{\mu\nu} \Bigl(\begin{smallmatrix} 0 & 0 \\[4pt] T^{\mu\nu} & 0 \end{smallmatrix}\Bigr) \,, \\[8pt]
V_R &= \Bigl(\begin{smallmatrix} \quad i\pdsl \quad & 0 \\[4pt]
            0 & i\pdsl -\Gsl^\T - \Vsl^\T \end{smallmatrix}\Bigr) \,,&\qquad
S_R &= \Bigl(\begin{smallmatrix} 0 & S^\dagger \\[4pt] 0 & 0 \end{smallmatrix}\Bigr) \,,&\qquad
T_R &= \sigma_{\mu\nu} \Bigl(\begin{smallmatrix} 0 & T^{\dagger\mu\nu} \\[4pt] 0 & 0 \end{smallmatrix}\Bigr) \,.
\end{alignat}
\end{subequations}
These components satisfy the following relations under the (extended) charge conjugation defined in \cref{eqn:CCdef}:
\begin{equation}\label{eqn:CCVST}
\overline{V}_{L/R} = V_{R/L} \,,\qquad
\overline{S}_{L/R} = S_{L/R} \,,\qquad
\overline{T}_{L/R} = T_{L/R} \,.
\end{equation}

With the decomposition in \cref{eqn:X0decompose}, we can expand the traces in \cref{eq:tr} into a set of terms, each being a product of the components
\begin{equation}
V_{L/R}\, \frac{1\mp\gamma^5}{2} \,,\qquad
S_{L/R}\, \frac{1\mp\gamma^5}{2} \,,\qquad
T_{L/R}\, \frac{1\mp\gamma^5}{2} \,.
\end{equation}
The matrix structures of these components, their chiralities, and charge conjugation properties lead to simplifications of the calculation:
\begin{itemize}
\item For the Dirac trace to be nonzero, each term must have an even power of $\gamma^\mu$ matrices in total. Given the structures of the traces in \cref{eq:tr}, this implies that only terms with an even power of $V_{L/R}$ will contribute.
\item The matrix structure of $S_{L/R}$ tells us that
\begin{equation}
S_L \left( \cdots V_{L/R} \cdots \right) S_L = S_R \left( \cdots V_{L/R} \cdots \right) S_R = 0 \,,
\label{eqn:SSmatrix}
\end{equation}
where $\left( \cdots V_{L/R} \cdots \right)$ does not contain any $S_{L/R}$ or $T_{L/R}$ factors. The same is true if we replace any of the $S_{L/R}$ in \cref{eqn:SSmatrix} with $T_{L/R}$.
\item The product of the chirality projection factors $\frac{1\mp\gamma^5}{2}$ will impose further selection rules.
\item Finally, one can make use of the charge conjugation properties in \cref{eqn:CCVST} to merge terms and simplify the result.
\end{itemize}

Now we apply these constraints to the case of this subsection, where $S, V_\mu \ne 0$ but $T_{\mu\nu}=0$. It is easy to see that $\btr_0$ does not contain any $S$-dependent terms, while the nonzero terms in $\btr_1, \btr_2, \btr_3$ must have two powers of $S$ and two powers of $V$ with appropriate chirality combinations. Starting with $\btr_1$, we get
\begin{equation}
\btr_1^{(S^2V^2)} = \frac12 \tr \Bigl[ \bigl( S_R V_L S_L V_R - V_L S_L V_R S_R + S_L V_R S_R V_L - V_R S_R V_L S_L \bigr)\, \alphamat\Bigr] \,,
\label{eq:tr1_S2V2}
\end{equation}
where terms containing one power of $\gamma^5$ have been dropped since $\tr (\gamma^\mu\gamma^\nu\gamma^5)=0$. We can use charge conjugation to further simplify this trace. Upon cyclic permutation the four terms in \cref{eq:tr1_S2V2} combine in pairs and give
\begin{align}
\btr_1^{(S^2V^2)} = \tr \Big[ \bigl( S_R V_L S_L V_R - V_R S_R V_L S_L \bigr)\, \alphamat \Big] = \tr \Big( S_R V_L S_L\, \comm{V_R}{\alphamat} \Big) \,.
\label{eq:tr1_S2V2_combine}
\end{align}
The other two traces $\btr_2$ and $\btr_3$ admit similar simplifications. The general rule we follow is to rewrite half of the terms using charge conjugation such that the entire expression is proportional to the commutator $\comm{V_R}{\alphamat}$. After contracting the gamma matrices using $\gamma^\mu\gamma_\mu = 4$, $\gamma^\mu \gamma^\nu \gamma_\mu = -2\gamma^\nu$, we find
\begin{subequations}
\begin{align}
\btr_2^{(S^2V^2)} &= \tr \Big\{ \bigl( S_R V_R S_L - 2 V_L S_R S_L - 2 S_R S_L V_L \bigr)\, \comm{V_R}{\alphamat} \Big\} \,, \\[8pt]
\btr_3^{(S^2V^2)} &= \tr \Big( S_R S_L\, \comm{V_R}{\alphamat} \Big)
= \tr \Big\{ \bigl( S_R S_L V_R + V_R S_R S_L \bigr)\, \comm{V_R}{\alphamat} \Big\} \,.
\end{align}
\end{subequations}

Combining the three traces above and substituting in the expressions for $S_{L,R}\,, V_{L,R}$ from \cref{eqn:chiralVST}, we find that the additional contribution to the anomaly from scalar couplings is
\begin{align}
\Acal_{\beta=0}^\Lambda[\alpha]\Bigr|_{\O(S^2V^2)} &= -\frac{1}{192\s\pi^2} \int \dd^4x\, \tr \bigg\{ \Big[
S^\dagger ( \Gsl^\T + \Vsl^\T ) S - S^\dagger S\, ( \Gsl + \Vsl )
\nonumber\\[5pt]
&\hspace{100pt}
- (\Gsl + \Vsl)\, S^\dagger S
+ i\bigl( S^\dagger \overleftrightarrow{\Dsl_V} S\bigr) \Big] (\pdsl \alpha) \bigg\} \,,
\label{eqn:AcalS2V2}
\end{align}
where $\bigl( S^\dagger \overleftrightarrow{\slashed{D}_V} S\bigr) \equiv \gamma_\mu \bigl[ S^\dagger (D_V^\mu S) - (D_V^\mu S^\dagger) S\bigr]$. Here we have defined a shifted covariant derivative $D_V^\mu$ that also contains the vector interactions $V^\mu$:
\begin{equation}
D_V^\mu \equiv D^\mu \bigr|_{G^\mu\to G^\mu+V^\mu} = \partial^\mu - i \left( G^\mu + V^\mu \right) \,.
\label{eqn:DVdef}
\end{equation}
Its action on $S, S^\dagger$ follows the same substitution:
\begin{equation}
\left(D_V^\mu S\right) \equiv \left(D^\mu S\right)\bigr|_{G^\mu\to G^\mu+V^\mu} \,, \qquad\qquad
\left(D_V^\mu S^\dagger\right) \equiv \left(D^\mu S^\dagger\right)\bigr|_{G^\mu\to G^\mu+V^\mu} \,.
\label{eq:DVS}
\end{equation}
If desired, one could easily evaluate the Dirac trace $\tr(\gamma^\mu\gamma^\nu) = 4\s\eta^{\mu\nu}$ in \cref{eqn:AcalS2V2}, but this is unnecessary for showing that it is an irrelevant anomaly.

To find the corresponding counterterm, we recall the gauge transformation of the scalar interactions $S[G_\mu, \phi]$ from \cref{eqn:GTvarious}:
\begin{equation}
S \quad\longrightarrow\quad
S_\alpha = U_\alpha^* S U_\alpha^\dagger \,,
\end{equation}
which leads to
\begin{equation}
S^\dagger S \longrightarrow U_\alpha S^\dagger S U_\alpha^\dagger \,,
\qquad\qquad
\delta_\alpha \left( S^\dagger S \right) = i\,\comm{\alpha}{S^\dagger S} \,.
\end{equation}
Their covariant derivatives by definition transform in the same way. This remains true for the shifted covariant derivative $D_V^\mu$ defined in \cref{eqn:DVdef}, and therefore we have
\begin{equation}
\delta_\alpha \big( S^\dagger \overleftrightarrow{\Dsl_V} S \big) = i\,\comm{\alpha}{\big( S^\dagger \overleftrightarrow{\Dsl_V} S \big)} \,.
\end{equation}
From the gauge transformation properties discussed above, together with those of $G_\mu$, $V_\mu$ in \cref{eqn:GTGV}, we can identify
\begin{equation}
\Acal_{\beta=0}^\Lambda[\alpha]\Bigr|_{\O(S^2V^2)} = -\delta_\alpha \int \dd^4x \,\Delta\L_\text{ct}^{(S^2V^2)} \,,
\end{equation}
where
\begin{align}
\Delta\L_\text{ct}^{(S^2V^2)} &= \frac{1}{192\pi^2} \int \dd^4x\, \tr \bigg[
\frac12\, S^\dagger (\Gsl^\T + \Vsl^\T )\, S\, (\Gsl + \Vsl )
\notag\\[5pt]
&\hspace{110pt}
- S^\dagger S (\Gsl + \Vsl) (\Gsl + \Vsl)
+ i\big( S^\dagger \overleftrightarrow{\Dsl_V} S \big)\, \Gsl \bigg] \,.
\label{eq:LctS2V2}
\end{align}
We therefore conclude that when both vector and scalar interactions are present, the additional contributions to the anomaly beyond the minimal coupling case are all irrelevant.

\subsection{Vector, Scalar, and Tensor Interactions}
\label{subsec:Tensor}

Finally, we also include the tensor interactions $T_{\mu\nu}$ alongside vector and scalar interactions in this subsection. The calculation proceeds in a similar way to the vector and scalar interactions case in the previous subsection; the gamma matrix algebra is slightly more tedious but it is straightforward.

Using the decomposition in \cref{eqn:X0decompose}, we immediately see that again, $\btr_0$ does not contain any $T_{\mu\nu}$-dependent terms. For $\btr_1, \btr_2, \btr_3$, the additional nonzero terms are of the form $TSV^2$ and $T^2V^2$. We examine them in turn below.

\paragraph{$TSV^2$ terms:}
Upon contraction of gamma matrices using $\gamma^\mu\gamma_\mu = 4$, $\gamma^\mu \gamma^\nu \gamma_\mu = -2\gamma^\nu$, and noting $\gamma^\mu T_{L,R} \gamma_\mu = 0$ (since $\gamma^\mu \gamma^\nu\gamma^\rho\gamma_\mu = 4\eta^{\nu\rho}$ while $T_{L,R}$ involves the antisymmetric $\sigma^{\mu\nu}$), we find
\begin{subequations}
\begin{align}
\btr_1^{(TSV^2)} &= \tr \Big\{ \big( T_R V_L S_L + S_R V_L T_L \big)\,
\comm{V_R}{\alphamat}\, (1+\gamma^5) \Big\} \,, \\[8pt]
\btr_2^{(TSV^2)} &= \tr \Big\{ \big( T_R V_R S_L + S_R V_R T_L 
\notag\\[3pt]
&\hspace{40pt}
- 2 V_L S_R T_L  - 2 T_R S_L V_L \big)\,
\comm{V_R}{\alphamat}\, (1+\gamma^5) \Big\} \,, \\[8pt]
\btr_3^{(TSV^2)} &= \tr \Big\{ T_R S_L\, \comm{V_R^2}{\alphamat}\, (1+\gamma^5)
+ S_R T_L\, \comm{V_R^2}{\alphamat}\, (1-\gamma^5) \Big\} \,.
\end{align}
\end{subequations}
To arrive at these equations we have combined terms that are related by charge conjugation and used cyclic permutation as in the previous subsection. We can further show that $\btr_3^{(TSV^2)}=0$ because
\begin{align}
\comm{V_R^2}{\alphamat} &= \Big( V_R^\mu\, \comm{V_R^\nu}{\alphamat} + \comm{V_R^\mu}{\alphamat}\, V_R^\nu \Big) \gamma_\mu \gamma_\nu
\notag\\[5pt]
&= \Big( \big[ V_R^\mu, \comm{V_R^\nu}{\alphamat} \big] + \comm{V_R^\nu}{\alphamat}\, V_R^\mu + \comm{V_R^\mu}{\alphamat}\, V_R^\nu \Big) \gamma_\mu \gamma_\nu \,.
\end{align}
The expression in parentheses is symmetric in $\mu \leftrightarrow \nu$ (note that for $\big[ V_R^\mu, \comm{V_R^\nu}{\alphamat}\big]$, only its upper-left block $(-\partial^\mu\partial^\nu\alpha)$ will eventually feed into the expressions), whereas the Dirac traces are antisymmetric:
\begin{subequations}
\begin{align}
\tr\bigl(\gamma_\mu \gamma_\nu \sigma_{\rho\tau} \bigr) &= -\tr\bigl(\gamma_\nu \gamma_\mu \sigma_{\rho\tau} \bigr) \,,\\[5pt]
\tr\bigl(\gamma_\mu \gamma_\nu \sigma_{\rho\tau} \gamma^5\bigr) &= -\tr\bigl(\gamma_\nu \gamma_\mu \sigma_{\rho\tau} \gamma^5\bigr) \,.
\end{align}
\end{subequations}
Adding up $\btr_1$ and $\btr_2$ and substituting in the expressions for $S_{L,R}$, $V_{L,R}$, $T_{L,R}$ from \cref{eqn:chiralVST}, we obtain
\begin{align}
\Acal_{\beta=0}^\Lambda[\alpha]\Bigr|_{\O(TSV^2)} &= - \frac{1}{192\pi^2} \int \dd^4x\, \tr \bigg\{ \Big[
(\sigma\cdot T^\dagger) ( \Gsl^\T+\Vsl^\T ) S + S^\dagger ( \Gsl^\T + \Vsl^\T ) (\sigma\cdot T)
\notag\\[5pt]
&\hspace{20pt}
+ 2i \Big( (\sigma\cdot T^\dagger) (\Dsl_V S ) - ( \Dsl_V S^\dagger) (\sigma\cdot T) \Big)
\Big] (\pdsl \alpha) (1+\gamma^5) \bigg\} \,,
\end{align}
where we have introduced the shorthand notation
\begin{equation}
\sigma\cdot T \equiv \sigma_{\mu\nu} T^{\mu\nu} \,,\qquad
\qquad
\sigma\cdot T^\dagger \equiv \sigma_{\mu\nu} T^{\dagger\mu\nu} \,.
\end{equation}
From the gauge transformation properties discussed earlier we see that
\begin{equation}
\Acal_{\beta=0}^\Lambda[\alpha]\Bigr|_{\O(TSV^2)} = -\delta_\alpha \int \dd^4x \,\Delta\L_\text{ct}^{(TSV^2)} 
\end{equation}
is an irrelevant anomaly corresponding to the following local counterterm:
\begin{align}
\Delta\L_\text{ct}^{(TSV^2)} &= \frac{1}{192\pi^2} \int \dd^4x\, \tr \bigg\{ \bigg[
\frac12 \Bigl( (\sigma\cdot T^\dagger) ( \Gsl^\T + \Vsl^\T ) S ( \Gsl + \Vsl ) 
\notag\\
&\hspace{140pt}
+ S^\dagger ( \Gsl^\T + \Vsl^\T ) (\sigma\cdot T) ( \Gsl + \Vsl ) \Bigr)
\notag\\[5pt]
&\hspace{65pt}
+ 2i\Big( (\sigma\cdot T^\dagger) (\Dsl_V S ) - ( \Dsl_V S^\dagger ) (\sigma\cdot T)
\Big) \Gsl \bigg] (1+\gamma^5) \bigg\} \,.
\label{eq:LctTSV2}
\end{align}

\paragraph{$T^2V^2$ terms:}
Finally, for the $T^2V^2$ terms, we find
\begin{subequations}
\begin{align}
\btr_1^{(T^2V^2)} &= \tr \Big\{T_R V_L T_L\, \comm{V_R}{\alphamat}\, (1+\gamma^5) \Big\} \,, \\[5pt]
\btr_2^{(T^2V^2)} &= \tr \Big\{ T_R V_R T_L\, \comm{V_R}{\alphamat}\, (1+\gamma^5) \Big\} \,, \\[5pt]
\btr_3^{(T^2V^2)} &= -\tr \Big\{ T_R \gamma_\mu T_L \gamma_\nu\, \comm{V_R^\mu V_R^\nu}{\alphamat}\, (1+\gamma^5) \Big\} \,,
\end{align}
\end{subequations}
where we have used $\gamma^\mu T_L V_R\gamma_\mu = 2\gamma_\mu T_L V_R^\mu$ to simplify $\btr_3$. Further, since the Dirac traces involved are symmetric under the exchange of $\gamma_\mu$ and $\gamma_\nu$:
\begin{subequations}
\begin{align}
\tr \bigl(\gamma_\mu \sigma_{\rho\tau} \gamma_\nu \sigma_{\kappa\lambda} \bigr) &= \tr \bigl(\gamma_\nu \sigma_{\rho\tau} \gamma_\mu \sigma_{\kappa\lambda} \bigr) \,,\\[5pt]
\tr \bigl(\gamma_\mu \sigma_{\rho\tau} \gamma_\nu \sigma_{\kappa\lambda} \gamma^5 \bigr) &= \tr \bigl(\gamma_\nu \sigma_{\rho\tau} \gamma_\mu \sigma_{\kappa\lambda} \gamma^5 \bigr) \,,
\end{align}
\end{subequations}
we can freely interchange $\mu$ and $\nu$ in $\btr_3$ and obtain
\begin{equation}
\btr_3^{(T^2V^2)} = - \tr \Big\{ \big(T_R \gamma_\mu T_L V_R^\mu + V_R^\mu T_R \gamma_\mu T_L \big)\, \comm{V_R}{\alphamat}\, (1+\gamma^5) \Big\} \,.
\end{equation}
Adding up all three traces and substituting in the expressions for $V_{L,R}\,, T_{L,R}$ from \cref{eqn:chiralVST}, we get
\begin{align}
\Acal_{\beta=0}^\Lambda[\alpha]\Bigr|_{\O(T^2V^2)} &= -\frac{1}{192\pi^2} \int \dd^4x\, \tr \bigg\{
\Big[ (\sigma\cdot T^\dagger) (\Gsl^\T + \Vsl^\T ) (\sigma\cdot T)
\notag\\[8pt]
&\hspace{30pt}
+ (\sigma\cdot T^\dagger) \gamma_\mu (\sigma\cdot T) ( G^\mu + V^\mu )
+ ( G^\mu + V^\mu ) (\sigma\cdot T^\dagger) \gamma_\mu (\sigma\cdot T)
\notag\\[5pt]
&\hspace{30pt}
+i \big( (\sigma\cdot T^\dagger) \overleftrightarrow{\Dsl_V} (\sigma\cdot T) \big)
\Big] (\pdsl \alpha) (1+\gamma^5) \bigg\} \,,
\end{align}\vspace{-1pt}
where $\big( (\sigma\cdot T^\dagger) \overleftrightarrow{\Dsl_V} (\sigma\cdot T) \big) =  \sigma_{\rho\tau} \gamma_\mu \sigma_{\kappa\lambda} \big[  T^{\dagger\rho\tau}  (D_V^\mu T^{\kappa\lambda}) - (D_V^\mu T^{\dagger\rho\tau}) T^{\kappa\lambda} \big]$.
This again can be identified with the gauge variation of a local counterterm:
\begin{equation}
\Acal_{\beta=0}^\Lambda[\alpha]\Bigr|_{\O(T^2V^2)} = -\delta_\alpha \int \dd^4x \,\Delta\L_\text{ct}^{(T^2V^2)} \,,
\end{equation}
where
\begin{align}
\Delta\L_\text{ct}^{(T^2V^2)} &= \frac{1}{192\pi^2} \int \dd^4x\, \tr \bigg\{ \bigg[ 
\frac12\, (\sigma\cdot T^\dagger) (\Gsl^\T + \Vsl^\T ) (\sigma\cdot T) ( \Gsl + \Vsl )
\notag\\[5pt]
&\hspace{120pt}
+ (\sigma\cdot T^\dagger) \gamma_\mu (\sigma\cdot T) \gamma_\nu ( G^\mu + V^\mu ) ( G^\nu + V^\nu )
\notag\\[5pt]
&\hspace{120pt} 
+ i\big( (\sigma\cdot T^\dagger) \overleftrightarrow{\Dsl_V} (\sigma\cdot T) \big) \Gsl
\bigg] (1+\gamma^5) \bigg\} \,.
\label{eq:LctT2V2}
\end{align}
We therefore conclude that additional contributions to $\Acal_{\beta=0}^\Lambda[\alpha]$ remain irrelevant when tensor couplings are included.

\subsection{Summary}
\label{subsec:Summary}

To summarize, in this section we have completed the calculation of the regularized anomaly $\Acal_\beta^\Lambda [\alpha]$ in the presence of scalar, vector, and tensor couplings to fermion bilinears and found that, with the Wess-Zumino consistent scheme choice $\beta=0$, the difference with respect to the minimal coupling case is an irrelevant anomaly:
\begin{equation}
\Acal_{\beta=0}^\Lambda[\alpha] = \Acal_{\beta=0}^{\Lambda,\text{mc}}[\alpha] - \delta_\alpha \int \dd^4x\, \Delta\L_\text{ct}^\Lambda \,.
\end{equation}
The corresponding local counterterm is
\begin{equation}
\Delta\L_\text{ct}^\Lambda \,=\, \L_{\text{ct},0}^\Lambda\Bigr|_{G_\mu\to G_\mu+V_\mu} - \L_{\text{ct},0}^\Lambda + \Delta\L_{\text{ct}}^{(V)} + \Delta\L_\text{ct}^{(S^2V^2)} + \Delta\L_\text{ct}^{(TSV^2)} + \Delta\L_\text{ct}^{(T^2V^2)} \,,\;\;
\end{equation}
with $\L_{\text{ct},0}^\Lambda$, $\Delta\L_{\text{ct}}^{(V)}$, $\Delta\L_\text{ct}^{(S^2V^2)}$, $\Delta\L_\text{ct}^{(TSV^2)}$ and $\Delta\L_\text{ct}^{(T^2V^2)}$ given by \cref{eq:Lct0,eq:LctV,eq:LctS2V2,eq:LctTSV2,eq:LctT2V2}, respectively. This means that for the renormalized anomaly $\Acal_\beta[\alpha]$ defined in \cref{eq:A_ren}:
\begin{equation}
\text{There exists a renormalization scheme where }\;
\Acal_{\beta=0}[\alpha] = \Acal_{\beta=0}^{\text{mc}}[\alpha] \,.
\end{equation}

\section{Discussion and Future Directions}
\label{sec:Discussion}

In this paper, we generalized the CDE framework for computing anomalies in Ref.~\cite{Paper1} to the case of relativistic EFTs with a general class of higher-dimensional operators. We systematically calculated the anomaly in this formalism, and demonstrated explicitly that the additional contributions from higher-dimensional operators are irrelevant anomalies. This means, in particular, that the (relevant) anomaly cancellation condition in SMEFT including the aforementioned higher-dimensional operators is the same as that in the Standard Model.

Our calculation did not include higher-dimensional operators which involve derivatives acting on the fermions (beyond the kinetic term), such as
\begin{equation}
\epsilon^{ik} \epsilon^{jl} \left( H_i D_\mu H_j \right) \left( \ell_k^\T i\gamma^0\gamma^2 D^\mu \ell_l \right) \,,\qquad
\left( H^\dagger D_\mu D_\nu H \right) \left( \bar\ell \gamma^\mu D^\nu \ell \right) \,.
\end{equation}
While there is no essential obstacle to incorporate them in our present formalism, the CDE calculation becomes more and more tedious with the inclusion of each derivative. Nevertheless, noting that the counterterms we found in \cref{eq:LctS2V2,eq:LctTSV2,eq:LctT2V2} share similar structures, we are hopeful that there could be a more efficient framework that would make such a calculation more manageable and potentially also shed new light on the underlying structures of CDE. We plan to pursue this intriguing possibility in future work.

The master functional trace evaluated in this paper, \cref{eq:anomaly_beta}, can also be relevant for certain EFT matching calculations, such as when integrating out heavy fermions that acquire masses from a Yukawa interaction via spontaneous symmetry breaking \cite{DHoker:1984izu,DHoker:1984mif}. Modern EFT matching calculations are typically performed with dimensional regularization. However, we anticipate our regularization prescription, applied in exclusively $d=4$ spacetime dimensions, should produce the same anomaly-related non-decoupling effects. We leave the exploration of this interesting question for future study.

\acknowledgments
\addcontentsline{toc}{section}{\protect\numberline{}Acknowledgments}

We thank Quentin Bonnefoy, Nathaniel Craig, Sungwoo Hong, Markus Luty and Aneesh Manohar for useful discussions. T.C.\ is supported by the U.S.\ Department of Energy under grant number DE-SC0011640. X.L.\ is supported by the U.S.\ Department of Energy under grant numbers DE-SC0009919 and DE-SC0011640. Z.Z.\ is supported by the U.S.\ Department of Energy under grant number DE-SC0011702. This work was performed in part at Aspen Center for Physics, which is supported by National Science Foundation grant PHY-1607611.

\end{spacing}

\begin{spacing}{1.09}
\addcontentsline{toc}{section}{\protect\numberline{}References}
\bibliographystyle{utphys}
\bibliography{Anomalies}
\end{spacing}

\end{document}